\documentstyle[aps,prb,manuscript,epsfig]{revtex}

\begin{document}

\def\epsfile{\epsfig}

\draft

\title{The generalized Kramers' theory for nonequilibrium open 
one-dimensional systems}

\author{Suman Kumar Banik, Jyotipratim Ray Chaudhuri and Deb Shankar Ray}

\address{Indian Association for the Cultivation of Science, Jadavpur,
Calcutta 700 032, India.}

\date{\today}

\maketitle

\begin{abstract}
The Kramers' theory of activated processes is generalized for nonequilibrium 
open one-dimensional systems. We consider both the internal noise due to thermal bath
and the external noise which are stationary, Gaussian and are characterized 
by arbitrary decaying correlation functions. We stress the role of a 
nonequilibrium stationary state distribution for this open system which
is reminiscent of an equilibrium Boltzmann distribution in calculation of 
rate. The generalized rate expression we derive here reduces to the 
specific limiting cases pertaining to the closed and open systems for thermal 
and non-thermal steady state activation processes.
\end{abstract}

\vspace{0.5cm}


\section{Introduction}

Ever since the seminal work of Kramers on the diffusion model of chemical 
reactions was published about half a century ago \cite{kram}, the theory
of activated processes has become a central issue in many areas of science
\cite{rmp,vim}, notably in chemical physics, nonlinear optics and condensed
matter physics. Kramers considered a model Brownian particle trapped in a 
one-dimensional well representing the reactant state which is separated by
a barrier of finite height from a deeper well signifying the product state.
The particle was supposed to be immersed in a medium such that the medium 
exerts a frictional force on the particle but at the same time  thermally 
activates it so that the particle may gain enough energy to cross the barrier.
Over several decades the model and many of its variants have served as 
standard paradigms in various problems of physical and chemical kinetics
to understand the rate in multidimensional systems in the overdamped and
underdamped limits \cite{rl,langer,ryter}, effect of anharmonicities 
\cite{ryter,ngv}, rate enhancement by
parametric fluctuations \cite{ph}, the role of non-Gaussian white noise
\cite{ngv,pgw1}, role of a relaxing bath \cite{jrc1,mm}, quantum and
semiclassical corrections 
\cite{pgw2,whm,aoc,ingold,grabert,weiss,jrc2}
to classical rate and related similar aspects. The
vast body of literature has been the subject of several reviews 
\cite{rmp,vim,ingold} and monograph \cite{weiss}.

The common feature of overwhelming majority of the aforesaid treatments is 
that
the system is thermodynamically closed which means that the noise of the 
medium is of {\it internal} origin so that the dissipation and fluctuations
get related through the fluctuation-dissipation relation \cite{kubo}. 
However, in a number
of situations the system is thermodynamically open, i.e., when the system is
driven by an {\it external} noise which is {\it independent} of system's
characteristic damping \cite{nit}. The distinctive feature of the dynamics in 
this case
is the absence of any fluctuation-dissipation relation. While in the former 
case a zero current steady state situation is characterized by an equilibrium
Boltzmann distribution, the corresponding situation in the latter case is 
defined only by a steady state condition, if attainable. It may therefore be 
anticipated \cite{vim}
that the independence of fluctuations and dissipation tends to
make the steady state distribution function depend on the strength and 
correlation time of external noise as well as on the dissipation of the 
system. The elucidation of the role of this steady state distribution in
rate theory is worth-pursuing.

Our aim in this paper is to generalize Kramers' theory of activated processes
for external noise in this context. We thus allow the Brownian particle in a 
potential field
to be driven by both external and internal stationary and Gaussian noise
fluctuations with arbitrary decaying correlation functions. The 
external noise may be of thermal or non-thermal type. We consider the
stochastic motion to be spatial-diffusion-limited and calculate the rate of 
escape over the barrier in the intermediate to strong damping regime within
an unified description. The
theory we develop here follows closely the original flux over population
method of Farkas \cite{farkas}. The
distinctive aspect, however, is the consideration of a steady state 
distribution instead of the equilibrium Boltzmann distribution for 
determination of quasi-stationary population in the source well. This affects 
the generalized rate expression significantly in two ways. First, the dynamics
around the bottom
of the source well exhibits the dependence of steady state distribution on the
dissipation. Second, the rate expression remains valid even in absence of any 
internal thermal noise. We mention, in passing, that the former point had 
earlier been rightly emphasized by Mel'nikov \cite{vim} as a specific 
requirement for a general theory.

Some pertinent points regarding the rate theory for nonequilibrium systems
may be in order. It is wellknown that though thermodynamically closed systems
with homogenous boundary conditions possess in general time-independent 
solutions, the driven or open systems may settle down to complicated multiple
steady states \cite{nit}
when one takes into account of nonlinearity of the system in 
full. Secondly in most nonequilibrium systems the lack of detailed balance
symmetry gives rise to severe problem in determination of stationary
probability density for multidimensional problem \cite{thomas}. 
Because of its one-dimensional
and linearized description the present treatment is free from these
difficulties. It is important to point out that the externally generated
nonequilibrium fluctuations can bias the Brownian motion of a particle
in an anisotropic medium and may used for design of molecular motors and
pumps \cite{science}. The nonequilibrium, non-thermal systems has also been 
investigated by a number of worker in different contexts, e.g., for examining
the role of colour noise in stationary probabilities \cite{paolo}, 
the properties of nonlinear systems \cite{moss}, 
the nature of cross-over \cite{jaume}, the effect of monochromatic noise 
\cite{ajmc},
the rate of diffusion-limited coagulation processes \cite{who}, etc.

The outlay of the paper is as follows: In Sec.II we generalize Kramers' theory 
of reaction rate for external noise. The stationary, Gaussian 
noise processes are of both external and internal type with arbitrary 
decaying correlation functions. A general form of steady state 
distribution function in the source well and a rate expression for 
barrier crossing dynamics for the nonequilibrium open system have been 
pointed out. In Sec.III we explicitly calculate the detailed form of the rate 
expressions for the specific cases. The paper is concluded in Sec.IV.

\section{Generalization of Kramers' theory for external noise}

We consider the motion of a particle of unit mass moving in a Kramers' type
potential $V(x)$ such that it is acted upon by random forces $f(t)$ and
$e(t)$ of both internal and external origin, respectively, in terms of the
following generalized Langevin equation
\begin{equation}
\label{eq1}
\ddot{x} + \int_0^t \gamma(t-\tau )\;  \dot{x}(\tau) \; d\tau + V'(x)
= f(t) + e(t) \; \; ,
\end{equation}

\noindent
where the friction kernel $\gamma(t)$ is connected to internal noise $f(t)$
by the wellknown fluctuation-dissipation relationship \cite{kubo}
\begin{equation}
\label{eq2}
\langle f(t) f(t') \rangle = k_B T \gamma(t-t') \; \; .
\end{equation}

\noindent
We assume that both the noises $f(t)$ and $e(t)$ are stationary and Gaussian. 
Their correlation times may be of arbitrary decaying type. The external noise 
is independent of the memory kernel and there is no corresponding 
fluctuation-dissipation relation. We further assume, without any loss of 
generality, that $f(t)$ is independent of $e(t)$ so that we have
\begin{equation}
\label{eq3}
\langle f(t) e(t) \rangle = 0 \; \; .
\end{equation}

The external noise modifies the dynamics of activation in two ways. First, it 
influences the dynamics in the region around the barrier top so that the
effective
stationary flux across it gets modified. Second, in presence of this noise
the equilibrium distribution of the source well is disturbed so that one has 
to consider a new stationary distribution, if any, instead of the standard 
Boltzmann
distribution. This new stationary distribution must be a solution of the
generalized Fokker-Planck equation around the bottom of the source well
region and serve as an appropriate boundary condition analogous to Kramers'
problem. We consider these two aspects separately in the next two
subsections.

\subsection{Fokker-Planck dynamics at the barrier top}

We consider the potential $V(x)$ as shown in Fig.1. Linearizing the potential 
around the barrier top at $x=0$ we write
\begin{equation}
\label{eq4}
V(x \approx 0) = V(0) - \frac{1}{2} \omega_b^2 x^2 + \ldots \; \; \; ;
\; \; \; \omega_b^2 > 0 \; \; .
\end{equation}

\noindent
Thus the Langevin equation takes the following form
\begin{equation}
\label{eq5}
\ddot{x} + \int_0^t \gamma(t-\tau )\;  \dot{x}(\tau) \; d\tau 
-\omega_b^2 x = F(t) 
\end{equation}

\noindent
where
\begin{equation}
\label{eq6}
F(t) = f(t) + e(t) \; \; .
\end{equation}

\noindent
The general solution of Eq.(\ref{eq5}) is given by,
\begin{equation}
\label{eq7}
x(t) = \langle x(t) \rangle + \int_0^t M_b(t-\tau)\; F(\tau) \; d\tau
\end{equation}

\noindent
where
\begin{equation}
\label{eq8}
\langle x(t) \rangle = v_0 M_b(t) + x_0 \chi_x^b (t)
\end{equation}

\noindent
with $x_0 = x(0)$ and $v_0 = \dot{x} (0)$ being the initial position and
velocity of the Brownian particle that are assumed to be nonrandom, and
\begin{equation}
\label{eq9}
\chi_x^b (t) = 1 + \omega_b^2 \int_0^t M_b (\tau) \; d\tau \; \; .
\end{equation}

\noindent
The kernel $M_b(t)$ is the Laplace inversion of,
\begin{equation}
\label{eq10}
{\tilde M}_b (s) = \frac{1}{s^2 + s {\tilde \gamma}(s) -\omega_b^2}
\end{equation}

\noindent
with
\begin{eqnarray*}
{\tilde \gamma}(s) = \int_0^\infty e^{-st} \; \gamma(t) \; dt \; \; .
\end{eqnarray*}

\noindent
The time derivative of Eq.(\ref{eq7}) gives
\begin{equation}
\label{eq11}
v(t) = \langle v(t) \rangle + \int_0^t m_b(t-\tau)\; F(\tau) \; d\tau
\end{equation}

\noindent
with
\begin{equation}
\label{eq12}
\langle v(t) \rangle = v_0 m_b(t) + \omega_b^2 x_0 M_b(t)
\end{equation}

\noindent
and
\begin{equation}
\label{eq13}
m_b (t) = \frac{d M_b (t)}{dt} \; \; .
\end{equation}

\noindent
Now using the symmetry of the correlation function,
\begin{eqnarray*}
\langle F(t) F(t') \rangle = C(t-t') = C(t'-t)
\end{eqnarray*}

\noindent
we compute the explicit expressions of the variances in terms of $M_b(t)$ and
$m_b(t)$ as,
\begin{mathletters}

\begin{eqnarray}
\sigma_{xx}^2 (t) & \equiv & \langle [ x(t) - \langle x(t) \rangle ]^2 
\rangle \nonumber \\
& = & 2 \int_0^t M_b(t_1) \; dt_1 \int_0^{t_1} M_b(t_2) \; C(t_1 - t_2) \; 
dt_2 \; \; ,\\
\sigma_{vv}^2 (t) & \equiv & \langle [ v(t) - \langle v(t) \rangle ]^2 
\rangle \nonumber \\
& = & 2 \int_0^t m_b(t_1) \; dt_1 \int_0^{t_1} m_b(t_2) \; C(t_1 - t_2) \;
dt_2 \; \; ,\\
\sigma_{xv}^2 (t) & \equiv & \langle [ x(t) - \langle x(t) \rangle ]
\; [ v(t) - \langle v(t) \rangle ] \rangle \nonumber \\
& = &  \int_0^t M_b(t_1) \; dt_1 \int_0^t m_b(t_2) \; C(t_1 - t_2) \; dt_2
\end{eqnarray}

\noindent
and from (14a) and (14c) we see that
\begin{equation}
\sigma_{xv}^2 (t) = \frac{1}{2} \dot{\sigma}_{xx}^2 (t) \; \; .
\end{equation}

\end{mathletters}

\noindent
While calculating the variances it should be remembered that by virtue of
Eq.(\ref{eq6})
\begin{equation}
\label{eq15}
C(t-t') = \langle f(t) f(t') \rangle + \langle e(t) e(t') \rangle \; \; .
\end{equation}

Since, in principle we know all the average quantities and variances of the
linear system driven by Gaussian noise one can make use of the characteristic 
function method to
write down the Fokker-Planck equation for phase space
distribution function $p(x,v,t)$ near the barrier top \cite{aad,mazo}
\begin{eqnarray}
\label{eq16}
\left [ \frac{\partial}{\partial t} + v \frac{\partial}{\partial x}
+ \bar{\omega}_b^2 (t) x \frac{\partial}{\partial v} \right ] p(x,v,t)
& = & \bar{\gamma}_b (t) \frac{\partial}{\partial v} v p(x,v,t) 
\nonumber \\
& & +
\phi_b(t) \frac{\partial^2}{\partial v^2} p(x,v,t) + \psi_b(t)
\frac{\partial^2}{\partial v \partial x} p(x,v,t)
\end{eqnarray}

\noindent
with
\begin{mathletters}

\begin{eqnarray}
\bar{\gamma}_b (t) & = & -\frac{d}{dt} \ln \Upsilon_b (t)  \; \; , \\
\bar{\omega}_b^2 (t) & = & 
\frac{-M_b (t) \dot{m}_b (t) + m_b^2 (t)}{\Upsilon_b (t)} \; \; ,\\
\Upsilon_b (t) & = & -\frac{m_b (t)}{\omega_b^2} \left \{ 1 + \omega_b^2
\int_0^t M_b (\tau) \; d\tau \right \} + M_b^2 (t) \; \; ,\\
\phi_b (t) & = & \bar{\omega}_b^2 (t) \sigma_{xv}^2 +
\bar{\gamma}_b (t) \sigma_{vv}^2 + \frac{1}{2} \dot{\sigma}_{vv}^2
\; \; {\rm and} \\
\psi_b (t) & = & \bar{\omega}_b^2 (t) \sigma_{xx}^2 +
\bar{\gamma}_b (t) \sigma_{xv}^2 + \dot{\sigma}_{xv}^2
- \sigma_{vv}^2
\end{eqnarray}

\end{mathletters}

\noindent
Regarding the Fokker-Planck equation (\ref{eq16}) three points are to be
noted. First, although bounded the time dependent functions 
$\bar{\gamma}_b(t)$, $\phi_b (t)$ and $\psi_b (t)$ may not always provide
long time limits. These play a decisive role in the calculation of
non-Markovian Kramers' rate. Therefore, in general, one has to work out
frequency $\bar{\omega}_b (t)$ and friction $\bar{\gamma}_b (t)$ functions
for analytically tractable models \cite{aad}.
Second, when the noise is purely internal (i.e., there exist a
fluctuation-dissipation relation) we have \cite{aad}
\begin{equation}
\label{eq18}
\phi_b (t)  =  k_B T \bar{\gamma}_b (t) \; \; {\rm and} \; \; 
\psi_b (t)  =  \frac{k_B T}{\omega_b^2} [ \bar{\omega}_b^2 (t) -
\omega_b^2 (t) ] \; \; .
\end{equation}

\noindent
Third, for pure external noise with Markovian relaxation, i.e., 
$\gamma (t) = \gamma \delta (t)$ we have
\begin{equation}
\label{eq19}
\bar{\gamma}_b (t) = \gamma \; , \; \bar{\omega}_b^2 (t) = \omega_b^2
\; , \; \phi_b (t) = \int_0^t C(t') \; m_b (t') \; dt' \; {\rm and} \;
\psi_b (t) = \int_0^t C(t') \; M_b (t') \; dt' \; \; .
\end{equation}

\subsection{Stationary distribution in the source well}

In order to calculate the stationary distribution near the bottom of the
left well we now linearize the potential $V(x)$ around $x=x_a$. The 
corresponding Fokker-Planck equation can be constructed using the 
above-mentioned technique to obtain
\begin{eqnarray}
\label{eq20}
\left [ \frac{\partial}{\partial t} + v \frac{\partial}{\partial x}
- \bar{\omega}_0^2 (t) x \frac{\partial}{\partial v} \right ] p(x,v,t)
& = & \bar{\gamma}_0 (t) \frac{\partial}{\partial v} v p(x,v,t) 
\nonumber \\
& & +
\phi_0(t) \frac{\partial^2}{\partial v^2} p(x,v,t) + \psi_0 (t)
\frac{\partial^2}{\partial v \partial x} p(x,v,t)
\end{eqnarray}

\noindent
with
\begin{mathletters}

\begin{eqnarray}
\bar{\gamma}_0 (t) & = & -\frac{d}{dt} \ln \Upsilon_0 (t)  \; \; , \\
\bar{\omega}_0^2 (t) & = & 
\frac{-M_0 (t) \dot{m}_0 (t) + m_0^2 (t)}{\Upsilon_0 (t)} \; \; ,\\
\Upsilon_0 (t) & = & -\frac{m_0 (t)}{\omega_0^2} \left \{ 1 - \omega_0^2
\int_0^t M_0 (\tau) \; d\tau \right \} + M_0^2 (t) \; \; ,\\
\phi_0 (t) & = & \bar{\omega}_0^2 (t) \sigma_{xv}^2 +
\bar{\gamma}_0 (t) \sigma_{vv}^2 + \frac{1}{2} \dot{\sigma}_{vv}^2
\; \; {\rm and} \\
\psi_0 (t) & = & \bar{\omega}_0^2 (t) \sigma_{xx}^2 +
\bar{\gamma}_0 (t) \sigma_{xv}^2 + \dot{\sigma}_{xv}^2
- \sigma_{vv}^2 \; \; .
\end{eqnarray}

\end{mathletters}

\noindent
Here the subscripts `0' signifies the dynamical quantities corresponding
to the bottom of the left well.

It may be easily checked that the stationary solution of Eq.(\ref{eq20})
is given by
\begin{equation}
\label{eq22}
p_{st}^0(x,v) = \frac{1}{Z} \exp \left [ -\frac{v^2}{2D_0} - 
\frac{ \tilde{V}(x)}{D_0 + \psi_0} \right ]
\end{equation}

\noindent
where, $D_0 = \phi_0/\bar{\gamma}_0$; $\psi_0$, $\phi_0$ and $\bar{\gamma}_0$
are the values at long time limit and $Z$ is the normalization constant.
Here $\tilde{V}(x)$ is the renormalized linearized potential with a 
renormalization in its frequency. 

It must be emphasized that the distribution (\ref{eq22}) is not an 
equilibrium distribution. This stationary distribution for the open system
plays the role of an equilibrium distribution for the closed system which 
may be however recovered in the absence of external noise terms. We also
point out in passing that because of the linearized potential $\tilde{V} (x)$
the steady state is unique and the question of multiple steady states does
not arise.

\subsection{Stationary current across the barrier}

In the spirit of Kramers' celebrated ansatz \cite{kram} we now demand a 
solution of the Eq.(\ref{eq16}) at the stationary limit of the type
\begin{equation}
\label{eq23}
p_{st}(x,v) = \exp \left [ -\frac{v^2}{2D_b} - 
\frac{ \tilde{V}(x)}{D_b + \psi_b} \right ] \; \xi (x,v)
\end{equation}

\noindent
with $D_b = \phi_b/\bar{\gamma}_b$ and $\psi_b$ are the long time limits of
the corresponding time dependent quantities specific for the barrier top
region. The notable difference from the Kramers ansatz is that the
exponential factor in (\ref{eq23}) is not the Boltzmann factor but pertains
to the dynamics at the barrier top.

The ansatz of the form (\ref{eq23}) denoting the steady state distribution
is motivated by the local analysis near the bottom and top of the barrier
in the Kramers' sense. For a nonequilibrium system, as in the present problem
of external time-dependent potential field, the relative population of the
two regions, in general, depends on the global properties of the potential.
Thus although at equilibrium the probability density is given by a Boltzmann
distribution, the external modulation of the potential requires energy input
and drives the system away from equilibrium, disturbing the Boltzmann
distribution. At this point one may anticipate the signature of dynamics in
the Kramers'-like ansatz (\ref{eq23}) compared to the standard Kramers'
ansatz for closed system (i.e., when the external field is absent). Thus
while in the latter case one considers a complete factorization of the
equilibrium part (Boltzmann) and the dynamical part, $\xi (x,v)$, the ansatz
(\ref{eq23}) incorporates the additional dynamical contribution through
dissipation and the strength of the noise into the exponential part. This 
explicit dynamical modification of Kramers' ansatz in the form of (\ref{eq23})
is valid so long as the extra dynamical contribution in the exponential factor
in (\ref{eq23}) does not become too severe, i.e., the amplitude of the 
external noise field is not too strong. To put it in a more quantitative way,
this implies (assuming for simplicity $D_0 \simeq D_b \sim D$,
$\psi_0 \simeq \psi_b \sim \psi$) that the thermal length scale, i.e., 
the maximum value of 
$\sqrt{D+\psi}/\gamma$ on which the velocity of the particle is thermalized,
should be shorter than the other characteristic length scales of the system, 
e.g., 
\begin{equation}
\label{eq71}
\sqrt{D+\psi}/\gamma \; < \; \sqrt{D/\omega_0^2} \; \left ( {\rm or} \;
\sqrt{D/\omega_b^2} \; \right ) \; \; .
\end{equation}

\noindent
These considerations are necessary for making spatial diffusion regime
and quasi-stationary condition meaningful in the present context.

Now inserting (\ref{eq23}) in (\ref{eq16}) in the steady state we get
\begin{equation}
\label{eq24}
-\left ( 1+ \frac{\psi_b}{D_b} \right ) v \frac{\partial \xi}{\partial x}
- \left [ \frac{D_b}{D_b+\psi_b} \bar{\omega}_b^2 x + \bar{\gamma}_b v
\right ] \frac{\partial \xi}{\partial v} + 
\phi_b \frac{\partial^2 \xi}{\partial v^2} +
\psi_b \frac{\partial^2 \xi}{\partial v \partial x} = 0 \; \; .
\end{equation}

\noindent
At this point we set
\begin{equation}
\label{eq25}
u = v + a x \; \; ,
\end{equation}

\noindent
and with the help of the transformation (\ref{eq25}), Eq.(\ref{eq24}) is 
reduced to the following form
\begin{equation}
\label{eq26}
( \phi_b + a \psi_b ) \frac{d^2 \xi}{d u^2}
- \left [ \frac{D_b}{D_b+\psi_b} \bar{\omega}_b^2 x + \left \{
\bar{\gamma}_b + a \left ( 1 + \frac{\psi_b}{D_b} \right ) \right \} v
\right ] \frac{d \xi}{d u} = 0 \; \; .
\end{equation}

\noindent
Now, let
\begin{equation}
\label{eq27}
\frac{D_b}{D_b+\psi_b} \bar{\omega}_b^2 x + \left \{ \bar{\gamma}_b +
a \left ( 1 + \frac{\psi_b}{D_b} \right ) \right \} v = -\lambda u
\end{equation}

\noindent
where $\lambda$ is a constant to be determined later.

\noindent
From (\ref{eq25}) and (\ref{eq27}) we have
\begin{equation}
\label{eq28}
a_\pm = -\frac{B}{2A} \pm \sqrt{ \frac{B^2}{4A^2} + \frac{C}{A} }
\end{equation}

\noindent
with
\begin{equation}
\label{eq29}
A = 1 + \frac{\psi_b}{D_b} \; \; , \; \; B = \bar{\gamma}_b \; \; {\rm and}
\; \; C = \frac{D_b}{D_b +\psi_b} \bar{\omega}_b^2 \; \; .
\end{equation}

\noindent
By virtue of the relation (\ref{eq27}), Eq.(\ref{eq26}) becomes
\begin{equation}
\label{eq30}
\frac{d^2 \xi}{d u^2} + \Lambda u 
\frac{d \xi}{d u} = 0
\end{equation}

\noindent
where
\begin{equation}
\label{eq31}
\Lambda = \frac{\lambda}{\phi_b + a \psi_b} \; \; .
\end{equation}

The general solution of the homogenous differential equation (\ref{eq30})
is
\begin{equation}
\label{eq32}
\xi (u) = F_2 \int_0^u \exp \left ( -\frac{1}{2} \Lambda u^2 \right )\; du
+ F_1 \; \; ,
\end{equation}

\noindent
where $F_1$ and $F_2$ are the constants of integration.

\noindent
The integral in the Eq.(\ref{eq32}) converges for $|u| \rightarrow \infty$
if only $\Lambda$ is positive. 
The positivity of $\Lambda$ depends on the sign of $a$; so by virtue of
Eqs.(\ref{eq25}) and (\ref{eq27}) we find that the negative root of $a$,
i.e., $a_-$ guarantees the positivity of $\Lambda$ since
\begin{equation}
\label{eq33}
-\lambda a = C \; \; .
\end{equation}

To determine the value of $F_1$ and $F_2$ we impose the first boundary 
condition on $\xi$
\begin{equation}
\label{eq34}
\xi (x,v) \longrightarrow 0 \; \; \; {\rm for} \; \; \; 
x \longrightarrow +\infty \; \; {\rm and} \; {\rm all} \; v \; \; .
\end{equation}

\noindent
This condition yields
\begin{equation}
\label{eq35}
F_1 = F_2 \left ( \frac{\pi}{2\Lambda} \right )^{1/2} \; \; .
\end{equation}

\noindent
Inserting (\ref{eq35}) into (\ref{eq32}) we have as usual
\begin{equation}
\label{eq36}
\xi (u) = F_2 \left [ \left ( \frac{\pi}{2\Lambda} \right )^{1/2} +
\int_0^u \exp \left ( -\frac{1}{2} \Lambda u^2 \right )\; du \right ] \; \; .
\end{equation}

Since we are to calculate the current around the barrier top, we expand the
renormalized potential $\tilde{V}(x)$ around $x \approx 0$
\begin{equation}
\label{eq37}
\tilde{V}(x) \simeq \tilde{V}(0) - \frac{1}{2} \bar{\omega}_b^2 x^2
\; \; .
\end{equation}

\noindent
Thus with the help of (\ref{eq36}) and (\ref{eq37}), Eq.(\ref{eq23})
becomes
\begin{equation}
\label{eq38}
p_{st} (x\approx 0,v) = F_2 \; e^{ -\frac{ \tilde{V}(0) }{D_b+\psi_b} } \;
\left [ \left ( \frac{\pi}{2\Lambda} \right )^{1/2} \;
e^{ -\frac{v^2}{2D_b} } + {\cal F} (x \approx 0,v) \; e^{ -\frac{v^2}{2D_b} }
\right ]
\end{equation}

\noindent
with
\begin{equation}
\label{eq39}
{\cal F} (x,v) = \int_0^u \exp \left ( -\frac{1}{2} \Lambda u^2 \right )\; du 
\; \; .
\end{equation}

Now defining the steady state current $j$ across the barrier by
\begin{equation}
\label{eq40}
j = \int_{-\infty}^{+\infty} v \; p_{st} (x \approx 0, v) \; dv
\end{equation}

\noindent
we have using Eq.(\ref{eq38})
\begin{equation}
\label{eq41}
j = F_2  D_b \frac{ \sqrt{2\pi} }{ ( \Lambda + D_b^{-1} )^{1/2} } \;
\exp \left [ -\frac{ \tilde{V}(0) }{D_b+\psi_b} \right ] \; \; .
\end{equation}

\subsection{Stationary population in the left well}

Having obtained the steady state current over the barrier top we now look for
the value of the undetermined constant $F_2$ in Eq.(\ref{eq41}) in terms of 
the population in the left well. We show that this may be done by matching
two appropriate {\it reduced} probability distributions at the bottom of the
left well.

To do so we return to the Eq.(\ref{eq23}) which describes the steady state
distribution at the barrier top. Again with the help of (\ref{eq36}) we
have
\begin{equation}
\label{eq42}
p_{st} (x,v) = F_2 \; \left [ \left ( \frac{\pi}{2\Lambda} \right )^{1/2}
+ \int_0^u \exp \left ( -\frac{1}{2} \Lambda u^2 \right ) \; du \right ]
\; \exp \left [ -\frac{v^2}{2D_b} - \frac{ \tilde{V}(x)}{D_b+\psi_b}
\right ] \; \; .
\end{equation}

We first note that, as $x\longrightarrow -\infty$, the pre-exponential
factor in $p_{st}(x,v)$ reduces to the following form
\begin{equation}
\label{eq44}
F_2 [\ldots ] = F_2 \; \left ( \frac{2\pi}{\Lambda} \right )^{1/2} \; \; .
\end{equation}

\noindent
We now define a reduced distribution function in $x$
\begin{equation}
\label{eq45}
\tilde{p}_{st} (x) = \int_{-\infty}^{+\infty} p_{st} (x,v) \; dv \; \; .
\end{equation}

\noindent
Hence we have from (\ref{eq44}) and (\ref{eq45})
\begin{equation}
\label{eq46}
\tilde{p}_{st} (x) = 2\pi F_2 \; \left ( \frac{D_b}{\Lambda} \right )^{1/2}
\; \exp \left [ - \frac{ \tilde{V}(x)}{D_b+\psi_b} \right ] \; \; .
\end{equation}

\noindent
Similarly we derive the reduced distribution function in the left well,
around $x \approx x_a$ using (\ref{eq22}) as
\begin{equation}
\label{eq47}
\tilde{p}_{st}^0 (x) = \frac{1}{Z} \sqrt{2\pi D_0} \; 
\exp \left [ - \frac{ \tilde{V}(x_a)}{D_0+\psi_0} \right ]
\end{equation}

\noindent
where we have employed, the expansion of $\tilde{V} (x)$ as
\begin{equation}
\label{eq48}
\tilde{V} (x) \simeq \tilde{V} (x_a) + \frac{1}{2} \bar{\omega}_0^2
(x-x_a)^2 \; \; \; , \; \; \; x \approx x_a
\end{equation}

\noindent
and $Z$ as the normalization constant.

At this juncture we impose the second boundary condition that, at 
$x = x_a$
the reduced distribution function (\ref{eq46}) must go over to stationary 
reduced distribution function (\ref{eq47}) at the bottom of the left well.
Thus we have 
\begin{equation}
\label{eq49}
\tilde{p}_{st}^0 (x = x_a) = \tilde{p}_{st} (x = x_a) \; \; .
\end{equation}

\noindent
The above condition is used to determine the undetermined constant $F_2$
in terms of the normalization constant $Z$ of Eq.(\ref{eq22})
\begin{equation}
\label{eq50}
F_2 = \frac{1}{Z} \; \left ( \frac{\Lambda}{2\pi} \right )^{1/2} \;
\left ( \frac{D_0}{D_b} \right )^{1/2} \;
\exp \left [ - \frac{ \tilde{V}(x_a) }{D_0+\psi_0} \right ] \; 
\exp \left [  
\frac{ \tilde{V}(0) -\frac{1}{2}\bar{\omega}_b^2 x_a^2 }{D_0+\psi_0} \right ]
\; \; .
\end{equation}

\noindent
Evaluating the normalization constant by explicitly using the relation
\begin{equation}
\label{eq51}
\int_{-\infty}^{+\infty} \int_{-\infty}^{+\infty} p_{st}^0 (x,v) \; dx \; dv
= 1
\end{equation}

\noindent
and then inserting its value in (\ref{eq50}) we obtain
\begin{equation}
\label{eq52}
F_2 = \frac{ \bar{\omega}_0 }{2\pi} \; 
\left ( \frac{\Lambda}{2\pi} \right )^{1/2} \;
\frac{1}{ D_b^{1/2} (D_0+\psi_0)^{1/2} } \; \exp \left [ 
\frac{ \tilde{V}(0) -\frac{1}{2}\bar{\omega}_b^2 x_a^2 }{D_0+\psi_0} \right ]
\; \; .
\end{equation}

\noindent
Making use of the relation 
$\tilde{V} (x_a) = \tilde{V} (0) - \frac{1}{2} \bar{\omega}_b^2 x_a^2$ in
(\ref{eq52}) and then the value of $F_2$ in Eq.(\ref{eq41}) we arrive at the 
expression for the normalized current or barrier crossing rate
\begin{equation}
\label{eq53}
k = \frac{ \bar{\omega}_0 }{2\pi} \;  \frac{D_b}{ (D_0+\psi_0 )^{1/2} } \;
\left ( \frac{\Lambda}{ 1 + \Lambda D_b } \right )^{1/2} \;
\exp \left [ - \frac{ E }{D_b+\psi_b} \right ] \; \; .
\end{equation}

\noindent
where the activation energy $E$ is defined as
\begin{eqnarray*}
E = \tilde{V} (0) - \tilde{V} (x_a) \; \; ,
\end{eqnarray*}

\noindent
as shown in Fig.1. Since the temperature due to internal thermal noise, the
strength of the external noise and damping constant are buried in the
parameters $D_0$, $D_b$, $\psi_0$, $\psi_b$ and $\Lambda$ the general
expression (\ref{eq53}) looks somewhat cumbersome. We note that the 
subscripts `0' and `b' in $D$ or $\psi$ refer to the well or the barrier top 
region, respectively. We discuss it in greater detail in the next section.

\section{Generalized Kramers' rate : Internal vs. external noise}

Eq.(\ref{eq53}) is the central result of this paper. This generalizes the 
Kramers' expression for rate of the activated processes for the nonequilibrium 
open systems. Both the internal and the external noises may be of
arbitrary long correlation time. It is important to note that the 
pre-exponential dynamical factors as well as the exponential factor are 
modified due to the openness of the system. The modification of the 
exponential factor is due to the fact that depending on the strength of the
external noise $e(t)$ the system settles down to a stationary distribution
which does not coincide with the usual equilibrium Boltzmann distribution.
The system therefore attains the steady state at a different `effective'
temperature. This aspect will be clarified in greater detail when we consider 
the limiting case in subsection D. In general, both the factors in the rate 
depend on the strength
of the noise, correlation time of fluctuations of both external and internal 
noise processes and dissipation. The rate is spatial-diffusion-limited and is
valid for intermediate to strong damping regime. 
This validity must be appreciated in the present context of driven system
in the sense that while on the one hand thermal length scale of the system
must be short compared to other characteristic length scales of the sytem
corresponding to the inequality (\ref{eq71}), dissipation should 
also obey the restriction that during one round trip of the particle in
phase space (in action, angle space) under purely deterministic motion
corresponding to (\ref{eq1}), the energy dissipated is greater than the 
thermal energy, i.e.,
\begin{equation}
\label{eq72}
\gamma I(E) \; > \; \sqrt{D+\psi}
\end{equation}

\noindent
where $I(E)$, the action, is equivalent to unperturbed energy $E$ in the
weak friction limit. Both the inequalities (\ref{eq71}) and
(\ref{eq72}) are therefore relevant for quantifying the 
spatial-diffusion-limited intermediate to strong damping regime. In what 
follows we shall be
concerned with several limiting situations to illustrate the general
result (\ref{eq53}) systematically for both thermal and non-thermal activated
processes.

\subsection{Internal white noise}

We first consider the case with no external noise and the internal thermal 
noise is $\delta$-correlated. To this end we set 
\begin{equation}
\label{eq54}
e(t)=0 \; \; \;  {\rm and} \; \; \; 
\langle f(t)f(t') \rangle = k_BT \gamma \delta (t-t') \; \; .
\end{equation}

\noindent
Making use of the abbreviations in Eqs.(17) and (21) it follows after some 
algebra that
\begin{eqnarray*}
\psi_0 = \psi_b = 0 \; , \; D_b = D_0 = k_BT \; , \; 
\Lambda = \frac{\lambda}{\gamma k_BT} \; , \\
\lambda = -(a_- + \gamma ) \; \; {\rm and} \; \;
a_\pm = -\frac{\gamma}{2} \pm \sqrt{ \frac{\gamma^2}{4} + \omega_b^2 } \; \; .
\end{eqnarray*}

\noindent
The above relations reduce the general expression (\ref{eq53}) to classical
expression for Kramers' rate \cite{kram}
\begin{equation}
\label{eq55}
k = \frac{\omega_0}{2\pi\omega_b} \; \left [ \left ( \frac{\gamma^2}{4} +
\omega_b^2 \right )^{1/2} - \frac{\gamma}{2} \right ] \; e^{-E/k_BT} \; \; .
\end{equation}

\subsection{Internal colour noise}

Next we consider the case with no external noise but the internal noise is
of Ornstein-Uhlenbeck type \cite{ou1,ou2}. Thus we have
\begin{equation}
\label{eq56}
e(t) = 0 \; \; \; {\rm and} \; \; \; 
\langle f(t) f(t') \rangle = \frac{ {\cal D} }{\tau_c} e^{-|t-t'|/\tau_c}
\; \; .
\end{equation}

\noindent
Here ${\cal D}$ denotes the strength while $\tau_c$ refers to the correlation
time of the noise. Again from Eqs.(17), (18) and (21) along with (\ref{eq56})
we derive the following relations
\begin{eqnarray*}
D_0 = D_b = k_BT \; \; , \\
\psi_0 = d_0 \; k_BT \; \; ; \; \;  1 + d_0 = \bar{\omega}_0^2/\omega_0^2 
\; \; , \\
\psi_b = d_b \; k_BT \; \; ; \; \;  1 + d_b = \bar{\omega}_b^2/\omega_b^2 
\; \; , \\
\lambda = -[ \bar{\gamma}_b + (1+d_b) a_-] \; \; {\rm and} \\ 
a_\pm = \frac{1}{1+d_b} \; \left [ -\frac{ \bar{\gamma}_b }{2} 
\pm \sqrt{ \frac{ \bar{\gamma}_b^2}{4} + \bar{\omega}_b^2 } \right ] \; \; .
\end{eqnarray*}

\noindent
and hence the rate becomes
\begin{equation}
\label{eq57}
k = \frac{\omega_0}{2\pi\omega_b} \; \left [ \left ( 
\frac{ \bar{\gamma}_b^2}{4} + \bar{\omega}_b^2 \right )^{1/2} 
- \frac{ \bar{\gamma}_b }{2} \right ] \; e^{-E/k_BT} \; \; .
\end{equation}

\noindent
whereby we recover the result of Grote-Hynes \cite{ghynes} and 
H\"anggi-Mojtabai \cite{hanggi} 
obtained several years ago.

\subsection{External colour noise}

Next we consider the case where the noise is completely due to of external
source and the external noise is of Ornstein-Uhlenbeck type \cite{ou1,ou2}
so that we set
\begin{equation}
\label{eq58}
f(t) = 0 \; \; \; {\rm and} \; \; \; 
\langle e(t) e(t') \rangle = \frac{ {\cal D} }{\tau_c} e^{-|t-t'|/\tau_c}
\; \; .
\end{equation}

\noindent
Note that since in this case the dissipation is independent of fluctuations 
we may
assume Markovian relaxation so that $\gamma (t) = \gamma \delta (t)$ (see
also Eqs.(18) and (19) ).

\noindent
The above condition (\ref{eq58}) when used in Eqs.(17), (19) and (21)
we obtain after some lengthy algebra
\begin{eqnarray*}
\phi_0 = \frac{ {\cal D} }{1+\gamma \tau_c + \omega_0^2 \tau_c^2} \; \; , 
\; \; 
\phi_b = \frac{ {\cal D} }{1+\gamma \tau_c - \omega_b^2 \tau_c^2}
\; \; ; \\
\psi_0 = \frac{ {\cal D}\; \tau_c }{1+\gamma \tau_c + \omega_0^2 \tau_c^2} 
\; \; , \; \; 
\psi_b = \frac{ {\cal D}\; \tau_c }{1+\gamma \tau_c - \omega_b^2 \tau_c^2}
\; \; ; \\
\lambda = - [ \gamma + (1+\gamma \tau_c ) a_- ]  \; \; {\rm and} \\
a_\pm = \frac{1}{1+\gamma \tau_c} \; \left [
- \frac{\gamma}{2} \pm \sqrt{ \frac{\gamma^2}{4} + \omega_b^2 } \right ]
\; \; .
\end{eqnarray*}

\noindent
and the rate becomes
\begin{equation}
\label{eq59}
k = \frac{\omega_0}{2\pi\omega_b} \left (
\frac{ 1+\gamma \tau_c + \omega_0^2 \tau_c^2 }{
1+\gamma \tau_c - \omega_b^2 \tau_c^2 } \right )^{1/2} \; 
\left [ \left ( \frac{ \gamma^2}{4} + \omega_b^2 \right )^{1/2} 
- \frac{ \gamma }{2} \right ] \; 
\exp \left [ - \frac{ \gamma ( 1+\gamma \tau_c - \omega_b^2 \tau_c^2 ) }{
{\cal D} ( 1 +\gamma \tau_c ) } E \right ] \; .
\end{equation}

\noindent
It is interesting to note that the expression (\ref{eq59}) denotes the 
external noise-induced barrier crossing rate which crucially depends on the 
strength ${\cal D}$ and correlation time $\tau_c$ of the coloured noise. The
absence of temperature and the appearance of dissipation $\gamma$ 
explicitly demonstrates the non-thermal origin of the noise processes as well
as the absence of fluctuation-dissipation relation.

\subsection{Internal and external white noise}

We finally consider both the internal and external noise to be 
$\delta$-correlated, i.e., 
\begin{equation}
\label{eq60}
\langle e(t)e(t') \rangle = 2 \alpha \delta (t-t') \; \; {\rm and}
\; \; \langle f(t)f(t') \rangle = \gamma k_BT \delta (t-t')
\end{equation}

\noindent
$\alpha$ being the strength of the external white noise. Hence, by 
virtue of (15), (17) and (21) we have
\begin{eqnarray*}
D_0 = D_b = k_BT + \frac{\alpha}{\gamma} \; \; , \; \; \psi_0 = \psi_b = 0
\; \; , \\
\lambda = -(a_- + \gamma ) \; \; {\rm and} \; \;
a_\pm = -\frac{\gamma}{2} \pm \sqrt{ \frac{\gamma^2}{4} + \omega_b^2 } \; \; .
\end{eqnarray*}

\noindent
Hence the rate becomes
\begin{equation}
\label{eq61}
k = \frac{\omega_0}{2\pi\omega_b} \; \left [ \left ( \frac{\gamma^2}{4} +
\omega_b^2 \right )^{1/2} - \frac{\gamma}{2} \right ] \; 
\exp \left [ \frac{E}{ k_BT + (\alpha/\gamma) } \right ] \; \; .
\end{equation}

\noindent
In the limit $\alpha \rightarrow 0$ we recover the Kramers original result
(\ref{eq55}) for pure internal white noise. We note here that 
$\alpha/(\gamma k_B)$ defines a new `effective' temperature due to external 
noise. The effective temperature which depends on the strength of the
external noise had been discussed earlier by Bravo et. al. \cite{sancho}
in a somewhat different context. We note that while in the latter
case the bath is driven by external fluctuations, the present treatment 
concerns the direct driving of the reaction co-ordinate by external noise.

\section{Conclusions}

In this paper we have generalized Kramers' theory of activated processes for
nonequilibrium open systems. The theory takes into account of both internal 
and external Gaussian noise fluctuations with arbitrary decaying correlation 
functions in an unified way. The treatment is valid for intermediate to 
strong damping regime for spatial diffusive processes.

The main conclusions of our study are summarized as follows;

\noindent
(i) We have shown that not only the motion at the barrier top is influenced
by the dynamics, it has 
an important role to play in establishing the stationary
state near the bottom of the source well for the open systems. Thus the
stationary distribution function in the well depends crucially on the
correlation time of the external noise processes as well as on damping. This
is distinctly a different situation (but analogous) as compared to an 
equilibrium Boltzmann distribution in the source well for standard Kramers'
theory for closed systems.

\noindent
(ii) Provided the long time limits of the moments for the stochastic processes
exist, the expression for Kramers' rate for barrier crossing for open
systems we derive here, is general.

\noindent
(iii) We have checked and examined the various limits of the generalized
rate expression to obtain Kramers' rate, its non-Markovian counterpart as
well as the other cases for specific external noise processes in presence
and absence of the internal noise.

\noindent
(iv) We have shown that a rate for barrier crossing dynamics induced by
purely non-thermal Gaussian noise can be derived as an interesting limiting 
case of the generalized rate expression.

We conclude by noting that since the validity of the rate expression derived 
in the paper depends on the existence of long time limit of the moments 
for the stochastic processes, the theory cannot be directly extended to,
say, fractal noise processes. These and the related noise processes remain
outside the scope of the present treatment. Suitable extension of the
Kramers' theory in this direction is worth-pursuing.

\acknowledgments
SKB is indebted to Council of Scientific and Industrial Research (C.S.I.R.),
Government of India for partial financial support.


\newpage

\begin{figure}
\caption{
A schematic plot of Kramers' type potential used in the text.
\protect
\label{fig1}}
\end{figure}

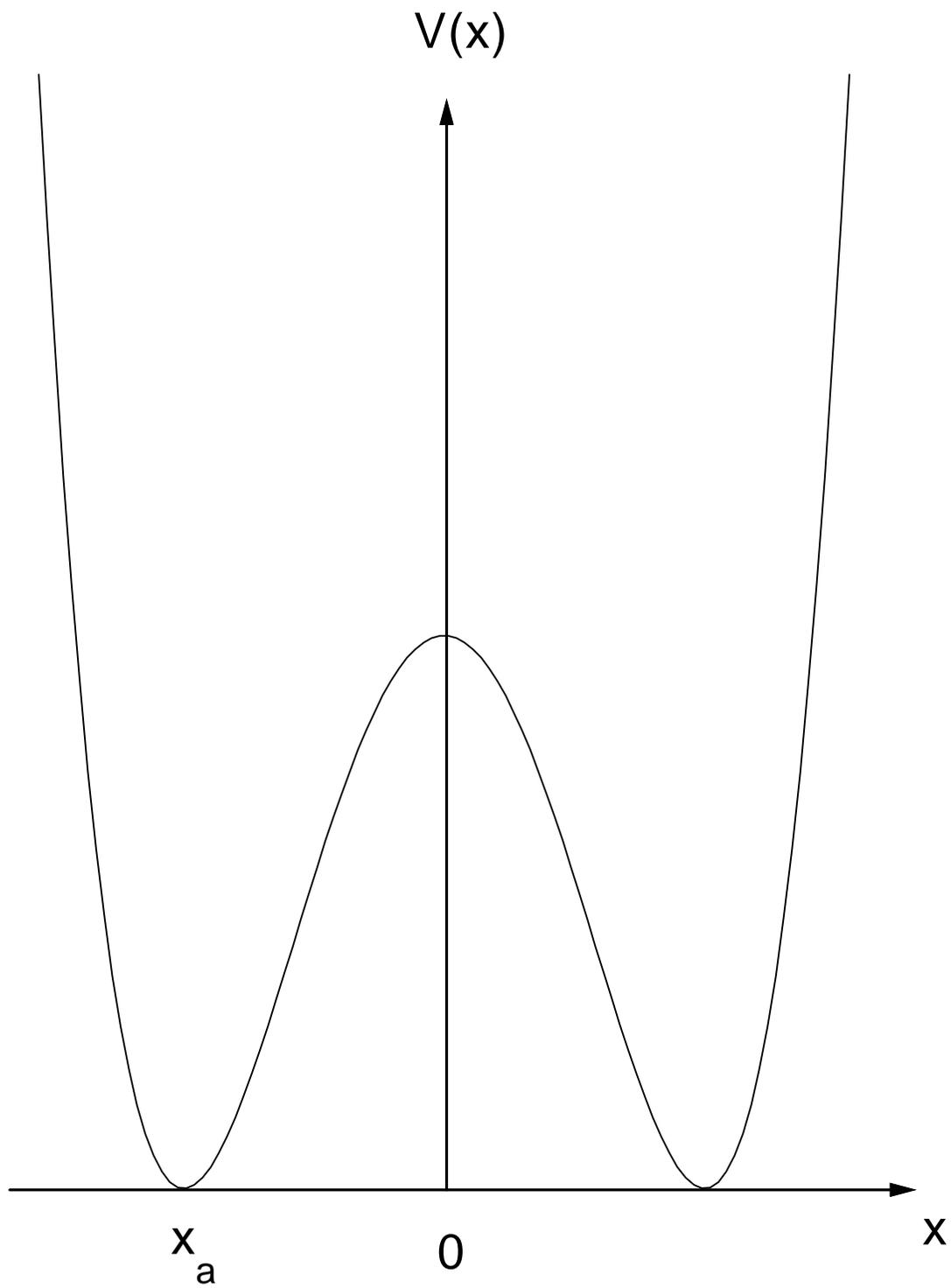
\end{document}